\documentclass[twocolumn,showpacs,aps,prl,superscriptaddress,floatfix]{revtex4}
\usepackage{graphicx}
\usepackage{epsfig}
\usepackage{amsmath}
\usepackage{amssymb}
\usepackage{hhline}
\usepackage{multirow}
\usepackage{rotating}
\usepackage{fancyheadings}
\usepackage{relsize}

\RequirePackage{xspace}

\def\babar{\mbox{\slshape B\kern-0.1em{\smaller A}\kern-0.1em
    B\kern-0.1em{\smaller A\kern-0.2em R}}}

\def\cbar  {\ensuremath{\overline c}\xspace}
\def\qbar  {\ensuremath{\overline q}\xspace}
\def\bbar  {\ensuremath{\overline b}\xspace}
\def\B       {\ensuremath{B}\xspace}
\def\Bbar    {\kern 0.18em\overline{\kern -0.18em B}{}\xspace}

\def\BB      {\ensuremath{B\Bbar}\xspace} 
\def\Bz      {\ensuremath{B^0}\xspace}
\def\Bzb     {\ensuremath{\Bbar^0}\xspace}
\def\BzBzb   {\ensuremath{\Bz {\kern -0.16em \Bzb}}\xspace}
\def\Bu      {\ensuremath{B^+}\xspace}
\def\Bub     {\ensuremath{B^-}\xspace}
\def\BpBm    {\ensuremath{\Bu {\kern -0.16em \Bub}}\xspace}
\def\jpsi     {\ensuremath{{J\mskip -3mu/\mskip -2mu\psi\mskip 2mu}}\xspace}
\def\psitwos  {\ensuremath{\psi{(2S)}}\xspace}
\def\Y#1S{\ensuremath{\Upsilon{(#1S)}}\xspace}
\def\FourS {\Y4S}
\def\Dbar    {\kern 0.2em\overline{\kern -0.2em D}{}\xspace}

\def\Dzb     {\ensuremath{\Dbar^0}\xspace}

\def\Dstar   {\ensuremath{D^*}\xspace}

\def\Dstarzb {\ensuremath{\Dbar^{*0}}\xspace}

\def\Dstarm  {\ensuremath{D^{*-}}\xspace}

\def\Ds      {\ensuremath{D^+_s}\xspace}
\def\piz     {\ensuremath{\pi^0}\xspace}
\def\epem    {\ensuremath{e^+e^-}\xspace}
\def\nue     {\ensuremath{\nu_e}\xspace}

\def\BR         {{\ensuremath{\cal B}\xspace}}

\newcommand{\gev}{\ensuremath{\mathrm{\,Ge\kern -0.1em V}}\xspace}
\newcommand{\mev}{\ensuremath{\mathrm{\,Me\kern -0.1em V}}\xspace}
\newcommand{\ev}{\ensuremath{\mathrm{\,e\kern -0.1em V}}\xspace}
\newcommand{\gevc}{\ensuremath{{\mathrm{\,Ge\kern -0.1em V\!/}c}}\xspace}
\newcommand{\mevc}{\ensuremath{{\mathrm{\,Me\kern -0.1em V\!/}c}}\xspace}
\newcommand{\gevcc}{\ensuremath{{\mathrm{\,Ge\kern -0.1em V\!/}c^2}}\xspace}
\newcommand{\mevcc}{\ensuremath{{\mathrm{\,Me\kern -0.1em V\!/}c^2}}\xspace}

\def\invfb   {\ensuremath{\mbox{\,fb}^{-1}}\xspace}

\newcommand{\stat}{\ensuremath{\mathrm{(stat)}}\xspace}
\newcommand{\syst}{\ensuremath{\mathrm{(syst)}}\xspace}

\def\pep2{PEP-II}

\newcommand{\nimBaseA}       {Nucl.\ Instr.\ Meth.\xspace}
\newcommand{\nim}       [1]  {\nimBaseA~{\bf #1}}
\def\mynim  #1 #2 #3 {\nim{#1},\ #2 (#3)}

\def\brpartbna{(9.66 \pm 0.27_{\stat} \pm 0.32_{\syst}) \%}
\def\brpartbca{(10.39 \pm 0.26_{\stat} \pm 0.37_{\syst})\%}
\def\brpartba{(10.03 \pm 0.19_{\stat} \pm 0.32_{\syst})\%}

\begin{document}

\begin{flushleft}                                      
\babar-PUB-06/36 \\                                    
SLAC-PUB-12027 \\                                        
\end{flushleft}                                        

\title[Short Title]{\large \bf Measurement of the Ratio ${\boldmath{\BR(\Bu \to X e \nu) / \BR(\Bz \to X e \nu)}}$}

%
\author{B.~Aubert}
\author{M.~Bona}
\author{D.~Boutigny}
\author{F.~Couderc}
\author{Y.~Karyotakis}
\author{J.~P.~Lees}
\author{V.~Poireau}
\author{V.~Tisserand}
\author{A.~Zghiche}
\affiliation{Laboratoire de Physique des Particules, IN2P3/CNRS et Universit\'e de Savoie,
 F-74941 Annecy-Le-Vieux, France }
\author{E.~Grauges}
\affiliation{Universitat de Barcelona, Facultat de Fisica, Departament ECM, E-08028 Barcelona, Spain }
% ...
\collaboration{The \babar\ Collaboration}
\noaffiliation

\begin{abstract}
We report measurements of the inclusive electron momentum spectra in decays
of charged and neutral B mesons, and of the ratio of semileptonic branching
fractions  ${\cal B}\xspace(\B^+ \to X e \nu)$ and ${\cal B}\xspace(\B^0 \to X e \nu)$. These were 
performed on  a sample of  231 million $B\kern 0.18em\overline{\kern -0.18em B}{}\xspace$ events  
recorded with the \mbox{\slshape B\kern-0.1em{\smaller A}\kern-0.1em B\kern-0.1em{\smaller A\kern-0.2em R}}
detector at the  $\Upsilon{(4S)}\xspace$ resonance. Events are  
selected by fully  reconstructing a hadronic  decay of  one $B$ meson  and  
identifying an electron among the decay products of the recoiling $\Bbar$ meson. We obtain 
${\cal B}\xspace(\Bu \to X e \nu)$/${\cal B}\xspace(\Bz \to X e \nu)$ = $1.084 \pm 0.041_{\mathrm{(stat)}}\xspace\pm 0.025_{\mathrm{(syst)}}\xspace$.
\end{abstract}


\maketitle

\def\bfbp{\ensuremath{\Bu \to X e \nu}}
\def\bfbn{\ensuremath{\Bz \to X e \nu}}
\def\bmesons{\B mesons}
\def\bbmeson{\Bbar meson}
\def\bmeson{\B meson}
\def\btag{\ensuremath{\B_{\rm tag}}}
\def\mes{\ensuremath{m_{ES}}}
\def\dsxf{\ensuremath{\Dstarm \leftrightarrow \Dstarzb \ }}
\def\rpm{\ensuremath{R_{+/0}}}

The hadronic decay widths of \Bu\ and \Bz\ mesons differ because of
mechanisms that depend on the flavor of the spectator quark, 
such as interactions involving the spectator quark or final state particles. 
This leads to different lifetimes $\tau_{\Bu}$ and $\tau_{\Bz}$ of charged
and neutral \B\ mesons. We do not expect different semileptonic decay
widths, since semileptonic decays do not involve the spectator quark. 
This means that the ratio $\rpm=\BR(\Bu \to X e \nu)$/$\BR(\Bz \to X e \nu)$
should agree with  $\tau_{\Bu}/\tau_{\Bz}$, which can be checked experimentally.

At the \FourS\ resonance, measurements of the inclusive semileptonic
branching fractions of \Bu\ and \Bz\ mesons are less precise than
for  an admixture  of $b$  hadrons.  The reason is mainly a  limitation of
statistics from the small efficiency of the event tag needed to separate
\BpBm\ from \BzBzb\ events. In this paper, we use fully reconstructed hadronic \B decays for this separation.
Combined with the high statistics of the \B factories, this approach allows
for a precision measurement of \rpm, as already demonstrated by the Belle
collaboration, measuring \rpm\  with 5\% uncertainty~\cite{belle}.
By tagging \BzBzb\ events with partially reconstructed $\Bz \to D^{*-} \ell \nu$ 
decays, the CLEO collaboration achieved a 14\% uncertainty on \rpm~\cite{Artuso}. 
High-momentum electron tags have been used
in similar analyses for the determination of $\BR(\B \to X e \nu)$ and
the electron momentum spectrum without separation of \Bz and \Bu 
decays~\cite{bbr1,bbr2}.

The measurements presented here are based on data collected  by the 
\babar\ detector~\cite{babar} at the PEP-II asymmetric $e^+e^-$ storage 
rings and correspond to an integrated luminosity of 
209~\invfb\ (231 million \BB\ events) on the \FourS\ resonance. For background and efficiency
corrections that cannot be measured directly from data, we use a full 
simulation of the detector based on GEANT4~\cite{geant}. The equivalent
luminosity of the simulated event sample amounts to about 980~\invfb 
for $\FourS \to \BB$ events and 300~\invfb for processes 
consisting of non-resonant $e^+e^- \to q\bar{q}$ $(q=u,d,s,c)$ 
production (``continuum events'').

In events with a fully reconstructed hadronic \B decay (\btag), we 
identify electrons among the remaining tracks. To avoid large backgrounds 
at lower momenta, we require $p_e > 0.6 \gevc$, where $p_e$ is the electron
momentum measured in the center-of-mass frame. Depending on the electron charge 
$q_e$ relative 
to the charge $q_b$ of the bottom quark in the \btag\  candidate, each electron is assigned
to either the right-sign ($q_e = - 3 q_b)$ or to the wrong-sign sample
($q_e = 3 q_b)$. In events without \BzBzb-mixing and a correctly reconstructed 
\btag, primary electrons from semileptonic decays of the signal \B 
are the dominant source for the right-sign sample, while 
electrons from $\B \to \Dbar X, \Dbar \to e^- \nue Y$ cascades 
populate the wrong-sign sample. 
We use the criteria in Ref.~\cite{bbr1} for track selection and electron 
identification, and apply the same procedures for efficiency and background
corrections of the right- and wrong-sign samples. In this analysis, 
we  additionally have to correct for mis-reconstructed \btag\ candidates.

Non-\BB events are suppressed by requiring the ratio of the second to the  
zeroth Fox-Wolfram moments~\cite{FoxWolfram} to be less than 0.5. To 
keep backgrounds from mis-reconstructed \btag\ candidates at a low level, 
we reconstruct hadronic \B decays in very pure modes only. To cancel 
systematic errors related to the \btag\  reconstruction, we select similar
(``twin'') modes for \Bz and \Bu decays~\cite{conjugate}:

\begin{table}[!h]
\begin{tabular}{lll}
(I) & $\B^{0} \to \pi (K \pi \pi)_{D^{-}}$ & 
$\B^{+} \to \pi(K \pi \pi^{0})_{\Dzb}$ \\

(II) & $\B^{0} \to \pi \left[(K \pi)_{\Dzb} \pi \right]_{\Dstarm}$ &  
$\B^{+} \to \pi \left[(K \pi)_{\Dzb} \pi^0 \right]_{\Dstarzb}$ \\

(III) & $\B^{0} \to \pi \pi \pi\left[(K \pi)_{\Dzb} \pi \right]_{\Dstarm}$ &
$\B^{+} \to \pi \pi \pi \left[(K \pi)_{\Dzb} \pi^0 \right]_{\Dstarzb}$ \\

(IV) & $\B^{0} \to \pi \left[(K \pi \pi^{0})_{\Dzb} \pi \right]_{\Dstarm}$ &
$\B^{+} \to \pi \left[(K \pi \pi^{0})_{\Dzb} \pi^0 \right]_{\Dstarzb}$ \\
(V) & $\B^{0} \to \pi \pi^{0}\left[(K \pi)_{\Dzb} \pi \right]_{\Dstarm}$ & 
$\B^{+} \to \pi \pi^{0}\left[(K \pi)_{\Dzb} \pi^0 \right]_{\Dstarzb}$ \\
\end{tabular}
\end{table}
Here $\pi$ and $K$ denote charged pions and kaons. The invariant mass of $\Dzb$ candidates is required to be within 15 \mevcc of
the nominal \Dzb mass~\cite{PDG} for the decay $\Dzb \to K \pi$ and 25 \mevcc for 
$\Dzb \to K \pi \piz$ decays. $D^-$ candidates are accepted if the invariant 
mass is within 20 \mevcc of the nominal $D^-$ mass. $D$ candidates with momenta 
above 2.5 \gevc (measured in the center-of-mass frame) are rejected since they 
indicate non-\BB events. 
$\Dstar$ candidates are built from pairs of $\Dzb$ 
candidates and charged (neutral) pions where the invariant mass difference 
$|M_{\Dzb\pi^{(0)}} - M_{\Dzb}|$ is within 2 \mevcc of the nominal mass difference.
In tag categories (III) and (V) we require the 
invariant masses $M_{\pi \pi \pi}$ and $M_{\pi \pi^{0}}$  to be less than 1.5 \gevcc. 
For further background reduction, we reject candidates where a kinematic fit with 
geometric constraints on the \B and $D$ vertices and mass constraints on the charmed 
mesons yields a $\chi^2$ value with a probability of less than 0.5\%. 

\begin{figure}[h]
\begin{center}
\includegraphics[width=3.4in]{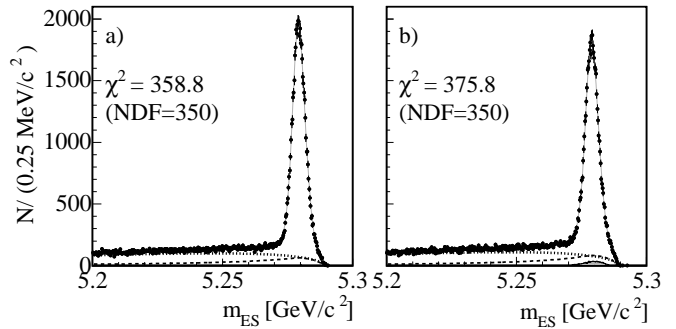} 
\caption{Fits of Eq.~\ref{eq:fsig} to distributions of the energy substituted mass for (a) neutral and (b) charged \btag\ candidates.
The dotted and dashed curves indicate the fitted contributions of continuum and 
combinatoric \btag\ candidates. The grey histogram displays the contribution of \dsxf\ background. }
\label{mestags}
\end{center}
\end{figure}

The kinematic consistency of the \btag\  candidates is checked with two variables, 
the beam-energy substituted mass $\mes=(s/4-p_B^2)^{1/2}$ and the energy 
difference $\Delta E = E_B - \sqrt{s}/2$. Here $\sqrt{s}$ refers to the total
center-of-mass energy, and $E_B$ and $p_B$ denote the energy and momentum of the \btag\ candidate,
all quantities being measured in the center-of-mass frame. For categories (I) - (III), we 
require $|\Delta E|< 50 \mev$, while the presence of an additional $\pi^0$ in (IV) and (V) 
leads to asymmetric distributions in $\Delta E$, motivating lower 
limits of $\Delta E>-75 \mev$ for (IV) and  $\Delta E>-100 \mev$ for (V). 
If for a given mode more than one \btag\ candidate satisfies these 
criteria, the one with the smallest $|\Delta E|$ is selected. Figure~\ref{mestags} 
shows the \mes\ distributions of \btag\ candidates 
satisfying these selection criteria. Candidates with 
$5.27 < \mes < 5.29 \gevcc$ are included in the \btag\ sample.
In $\approx 1\%$ of all events, we find
multiple \btag\ candidates in different decay modes. Here we use 
all of them, correcting for the background \btag\ candidates later.

The \btag\ sample can be divided into 4 components: signal, combinatoric 
background, \dsxf cross feed and continuum background. Correctly reconstructed 
\B decays are called {\it signal} \btag\ candidates, while \btag\ candidates
that contain tracks from the decay of the other \B contribute to the 
{\it combinatoric} \btag\ background. A special case of combinatoric background,
called \dsxf cross feed, contains cross feeds between twin modes of channels 
(II)-(V) due to mis-reconstruction of a \Dstarm as a \Dstarzb or vice versa. 
Due to the low energy of the combinatoric pion, the \mes\ distribution of this 
background is similar to the signal and will be treated separately from the
other combinatoric \btag\ background. The fourth component consists of 
\btag\ candidates arising from continuum events and is called {\it continuum} 
\btag\ background.  Since the ratio of signal to background \btag\ candidates 
depends on the multiplicity of the event and thus on the presence of a semileptonic 
decay, a precise determination of the number of signal \btag\ candidates is crucial 
to avoid biases in the branching fraction measurement. Monte Carlo (MC) studies using 
generator information indicate that once the \btag, right- and wrong-sign samples 
have been corrected for \btag\ background, the biases on the branching fraction 
measurements are below the statistical sensitivity given by the size of the MC sample, 
i.e. less than 0.5\%.

\renewcommand{\arraystretch}{1.5}
The contributions of combinatoric and continuum \btag\ background to the
\btag\ sample are extrapolated from the \mes\ sideband region, 
$5.2 < \mes < 5.25 \gevcc$. This requires a model of the background 
\mes\ distributions over the full range, $5.2 < \mes < 5.29 \gevcc$, which is 
obtained by fitting a linear combination of three functions describing the shapes of 
\mes\ distributions of signal, combinatoric and continuum \btag\ candidates 
to the observed \mes\ distributions. 

The shape of the combinatoric \btag\ background 
$f_{b\bbar}(\mes)$ is taken from the MC simulation. For the continuum 
background, we use the following function~\cite{ArgusFcn}:
$$f_{q\qbar}(m) =  m \ \sqrt{1-m^2} e^{-\kappa(1-m^2)} , \ $$
where $m=\mes/\mes^{\rm max}$ and $\mes^{\rm max}$ is the endpoint of the 
\mes\ distribution. 

For a given \B decay mode, the signal \mes\ distribution
is commonly described by a gaussian and a power law~\cite{CBallFcn}. Since
the \btag\ signal consists of many individual decay modes, a single function of
that type fails to describe our \mes\ distribution. We have found that a 
more general {\it ansatz} using a gaussian shape $f_g(x) = e^{-x^2/2}$ and
a function with a similar shape near $x=0$, but behaving like $e^{-x}$ for $x\to \pm \infty$,
$f_t(x) = e^{-x}/(1+e^{-x})^2$,  yields a good description of our 
signal \mes\ shape:
\begin{equation}
f_{\rm sig}(\Delta) =  \left\{
\begin{array}{lll}
\frac{C_2}{(C_3 - \Delta)^n}   & \text{if }  \Delta <   \alpha & \\
\frac{C_1}{\sigma_L} f_t(\frac{\Delta}{\sigma_L}) &   \text{if } \alpha \leq \Delta < 0 & \\
\frac{r}{\sigma_{1}} f_{t}\left(\frac{\Delta}{\sigma_{1}}\right) & + \ \frac{1-r}{\sigma_{2}} f_{g}\left(\frac{\Delta}{\sigma_{2}}\right) & \text{if } \Delta \geq 0\\ 
\end{array}
\right.  \ ,
\label{eq:fsig}
\end{equation}
with $\Delta=\mes - \overline{m}_{ES}$ and $\overline{m}_{ES}$ being the maximum of the
\mes\ distribution. $C_1$, 
$C_2$ and $C_3$ are functions of the parameters $\overline{m}_{\rm ES}$, $r$, $\sigma_1$, 
$\sigma_2$, $\sigma_L$, $\alpha$ and $n$ to ensure that $f_{\rm sig}$ is continuous 
and differentiable at $\Delta=0$ and $\Delta=\alpha$. 
This function, similar to the one featured in~\cite{CBallFcn}, describes the tails 
caused by the asymmetric energy resolution of neutral pions by a power law 
of order $-n$ and a junction $\alpha<0$ where it turns
into a gaussian-like shape. Fixing $\alpha$ and $n$ to the values obtained
from a fit to MC-simulated \mes\ distributions of signal \btag\ candidates, 
we fit a linear combination of $f_{q\qbar}$, $f_{b\bbar}$ and $f_{\rm sig}$ to the \mes\ distributions
observed in data, leaving all other parameters and normalizations free in
the fit (Fig.~\ref{mestags}). Due to their similar \mes\ distributions, this method cannot distinguish
between signal \btag\ candidates and \dsxf\ cross feed. This background contribution
is estimated from the MC simulation to be 0.5\% (2.6\%) relative to the signal for the neutral (charged) 
\btag\ sample.
\renewcommand{\arraystretch}{1}

To validate this extraction method, we perform the same analysis on our Monte 
Carlo sample and find that it reproduces the original number of signal 
\btag\ candidates. Uncertainties related to the MC simulation
of the combinatoric \btag\ background are evaluated by decomposing this background into 
the true underlying individual exclusive decay modes, and varying their 
contributions by the uncertainties of their branching fractions if they 
are reported in~\cite{PDG}, or $\pm$100\% otherwise. This leads to an uncertainty of 
$1.3\%$ on the number of $\Bz$ and $\Bu$ tags. Due to the different 
compositions of the combinatoric $\Bz$ and $\Bu$ backgrounds, these
errors are uncorrelated. In contrast, systematic errors related
to the description of the signal shape are correlated since we use
similar decay modes. Here we assess the uncertainties related to
the modeling of the shape for $\mes < \overline{m}_{ES}$ by repeating
the fit with $\alpha$ set to -$\infty$, allowing an
exponential function only instead of a power law to describe the
tail caused by the $\pi^0$ energy resolution. This leads to relative 
uncertainties of 2.1\% (2.4\%) on the number of \Bz (\Bu) tags.	
The yields of events in which \btag\ candidates have been found for both 
``twins'' of decay channels (II)-(V) differ by 20\% in data and MC, motivating
a relative uncertainty of 20\% on the \dsxf cross-feed.
This adds another systematic uncertainty of 0.5\% to the number
of charged \btag\ candidates. The final numbers of neutral and charged 
signal \btag\ candidates are
$ N_{\Bz}= 45420 \pm 420_{\stat} \pm 591_{(u)} \pm {949}_{(c)} $
and
$N_{\Bu}= 41948 \pm 463_{\stat} \pm 596_{(u)} \pm {1020}_{(c)} $, where
$u$ and $c$ denote uncorrelated and correlated systematic uncertainties respectively.

\begin{figure}[h]
\begin{center}
\includegraphics[width=3.4in]{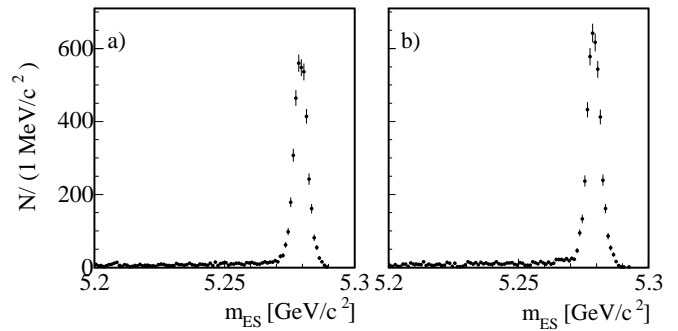} 
\caption{\mes\  distributions for (a) neutral and (b) charged \btag\ candidates in 
events with a right-sign electron.}
\label{mesrightsign}
\end{center}
\end{figure}

The requirement of an identified electron leads to significantly lower 
\btag\ backgrounds, as shown in Fig.~\ref{mesrightsign} for the 
right-sign sample. For high electron momenta ($p_e>1 \gevc$), the purities are 
96\% (98\%) for the right-sign (wrong-sign) samples, with combinatoric 
\btag\ candidates being the dominant background, while for decreasing
electron momenta, the purities decrease to 90\% because of an increasing 
amount of continuum-background. As for the full \btag\ sample, we estimate these backgrounds
from the \mes\ sideband region. The background estimates are performed separately for 
each sample as functions of $p_e$. Due to low
statistics, we do not determine the extrapolation factor from a fit, but 
use the MC predictions instead. The systematic errors due to 
the shape of the combinatoric background and \dsxf cross feed are evaluated 
in the same way as for the \btag\ sample, and the uncertainty in the continuum
contribution is taken to be 20\%.

\begin{figure}[ht]
\begin{center}
\includegraphics[height=3in]{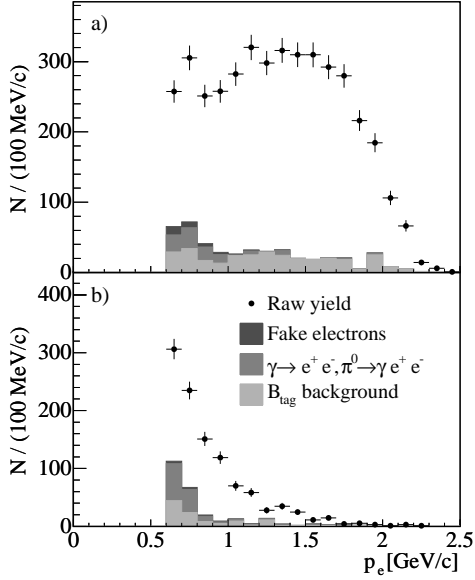} 
\caption{Total measured spectrum (points) and estimated backgrounds (histograms)
for electron candidates in events with a charged \btag\ candidate, for (a) the right-sign sample, 
and (b) the wrong-sign sample.}
\label{rawyields}
\end{center}
\end{figure}

Figure~\ref{rawyields} shows the momentum spectra of right- and wrong-sign 
electrons in events with a charged \btag\ candidate, together with the estimated
\btag\ background. This figure also displays the background contributions of 
electrons from photon conversions, $\piz \to \gamma \epem$ Dalitz decays and  
misidentified hadrons. These backgrounds are identified and corrected for as in~\cite{bbr1,bbr2}.
Corrections for electron identification efficiency and the evaluation of 
its systematic uncertainty are also performed as in~\cite{bbr1,bbr2}.

Background contributions from decays of charmed mesons produced in 
$b \to c\cbar s$ decays or decays of $\tau$ leptons are estimated from the
MC simulation, using the ISWG2 model~\cite{isgw2} to describe semileptonic
$D$ and $D_s$ meson decays. Assuming $\Gamma(D_s\to Xe\nu)=\Gamma(D\to Xe\nu)$, 
we obtain $\BR(D_s \to X e \nu) = (8.05 \pm 0.66)\%$. Inclusive $D_s$ production
has been measured in~\cite{bbrcharm} separately for neutral and charged \B decays, and
with~\cite{dsphipi} we obtain $\BR(\Bz \to \Ds \to e^+)$ = $(0.67 \pm 0.17)\%$ and
 $\BR(\Bu \to \Ds \to e^+)$ = $(0.88 \pm 0.18)\%$. Combining the measurements of
inclusive $D^{0}$ and $D^+$ production from~\cite{bbrcharm} with the inclusive 
$D^{0,+}\to e$ branching fractions from~\cite{PDG} yields $\BR(\Bz \to D^{+,0} \to e^+)$
=$(0.83 \pm 0.25)\%$ and  $\BR(\Bu \to D^{+,0} \to e^+)$ = $(1.33 \pm 0.20)\%$. Since
there are no branching fraction measurements for $\B \to \tau$ decays that distinguish
between neutral and charged \B decays, we assume $\Gamma(\Bz \to X \tau \nu)$ = 
$\Gamma(\Bu \to X \tau \nu)$ and combine the average value from~\cite{PDG} with the 
\B-meson lifetimes from direct measurements~\cite{PDG}. Including $\tau$ leptons
that originate from $\B \to D_s \to \tau$ cascades, we arrive at
$\BR(\Bz \to \tau \to e^+)$ =$(0.53 \pm 0.06)\%$ and  $\BR(\Bu \to \tau \to e^+)$ 
= $(0.60 \pm 0.06)\%$. Since the branching fractions of \B decays to $\jpsi$ and $\psitwos$
mesons are small and well measured, we use the MC simulation to correct for 
background electrons from $\jpsi \to \epem$ and $\psitwos \to \epem$ decays, using 
$\BR(\B \to \jpsi \to \epem)$ = $(6.49 \pm 0.22) \times 10^{-4}$ and 
$\BR(\B \to \psitwos \to \epem)$ = $(0.23 \pm 0.02) \times 10^{-4}$~\cite{PDG}. 

\begin{table*}[!ht]
\begin{center}
\caption{Electron yields for the four samples and corrections with statistical and systematic errors.}
\begin{tabular}{lp{2.5cm}p{2.5cm}p{2.5cm}p{2.5cm}}
\hline \hline
  &\Bz tags,\newline right-sign &\Bz tags,\newline  wrong-sign &\Bu tags,\newline  right-sign &\Bu tags,\newline  wrong-sign  \\
\hline 
$5.27 < \mes(\btag) < 5.29$ \gevcc  &$ 3461 \pm 59$ &$ 1943 \pm 44$ &$ 4074 \pm 64$ &$ 1070 \pm 33$  \\
$B_{tag}$ background  &$ 198 \pm 16 \pm 40$ &$ 135 \pm 13 \pm 27$ &$ 320 \pm 24 \pm 64$ &$ 114 \pm 12 \pm 23$  \\
$ \gamma \to \epem $ &$ 55 \pm 14 \pm 8$ &$ 87 \pm 17 \pm 12$ &$ 66 \pm 14 \pm 10$ &$ 83 \pm 16 \pm 11$  \\
$ \piz \to \gamma  \epem $ &$ 31 \pm 14 \pm 7$ &$ 25 \pm 12 \pm 5$ &$ 36 \pm 14 \pm 7$ &$ 47 \pm 16 \pm 9$  \\
fake $e$ &$ 29 \pm 1 \pm 8$ &$ 21 \pm 1 \pm 4$ &$ 37 \pm 1 \pm 12$ &$ 16 \pm 0 \pm 2$  \\
\hline 
Yield before and &$ 3149 \pm 64 \pm 42$ &$ 1674 \pm 51 \pm 30$ &$ 3616 \pm 71 \pm 66$ &$ 810 \pm 41 \pm 27$  \\
after $e$ efficiency correction &$ 3443 \pm 70 \pm 71$ &$ 1842 \pm 56 \pm 50$ &$ 3947 \pm 78 \pm 96$ &$ 898 \pm 46 \pm 41$  \\
\hline 
$\B \to (D_s \to) \tau \to e$ &$ 92 \pm 9 \pm 8$ &$ 20 \pm 4 \pm 2$ &$ 109 \pm 10 \pm 9$ &  0\\
$\B \to D_s \to e$ &$ 65 \pm 9 \pm 16$ &$ 13 \pm 4 \pm 3$ &$ 96 \pm 11 \pm 20$ &  0\\
$\B \to D \to e$ &$ 61 \pm 8 \pm 25$ &$ 12 \pm 4 \pm 5$ &$ 96 \pm 11 \pm 15$ &  0\\
$\B \to \jpsi,\psitwos \to e$ &$ 22 \pm 5 \pm 1$ &$ 23 \pm 5 \pm 1$ &$ 17 \pm 4 \pm 1$ &$ 18 \pm 4 \pm 1$  \\
\dsxf cross feed &$ 9 \pm 3 \pm 5$ &$ 4 \pm 2 \pm 2$ &$ 44 \pm 7 \pm 22$ &$ 29 \pm 5 \pm 15$  \\
\hline 
Net $e$ yield &$ 3195 \pm 72 \pm 82$ &$ 1769 \pm 57 \pm 51$ &$ 3585 \pm 81 \pm 106$ &$ 850 \pm 47 \pm 45$  \\
\hline \hline
\end{tabular} 
\label{tblyields}	
\end{center}
\end{table*}

\begin{figure}[t]
\begin{center}
\includegraphics[width=3.4in]{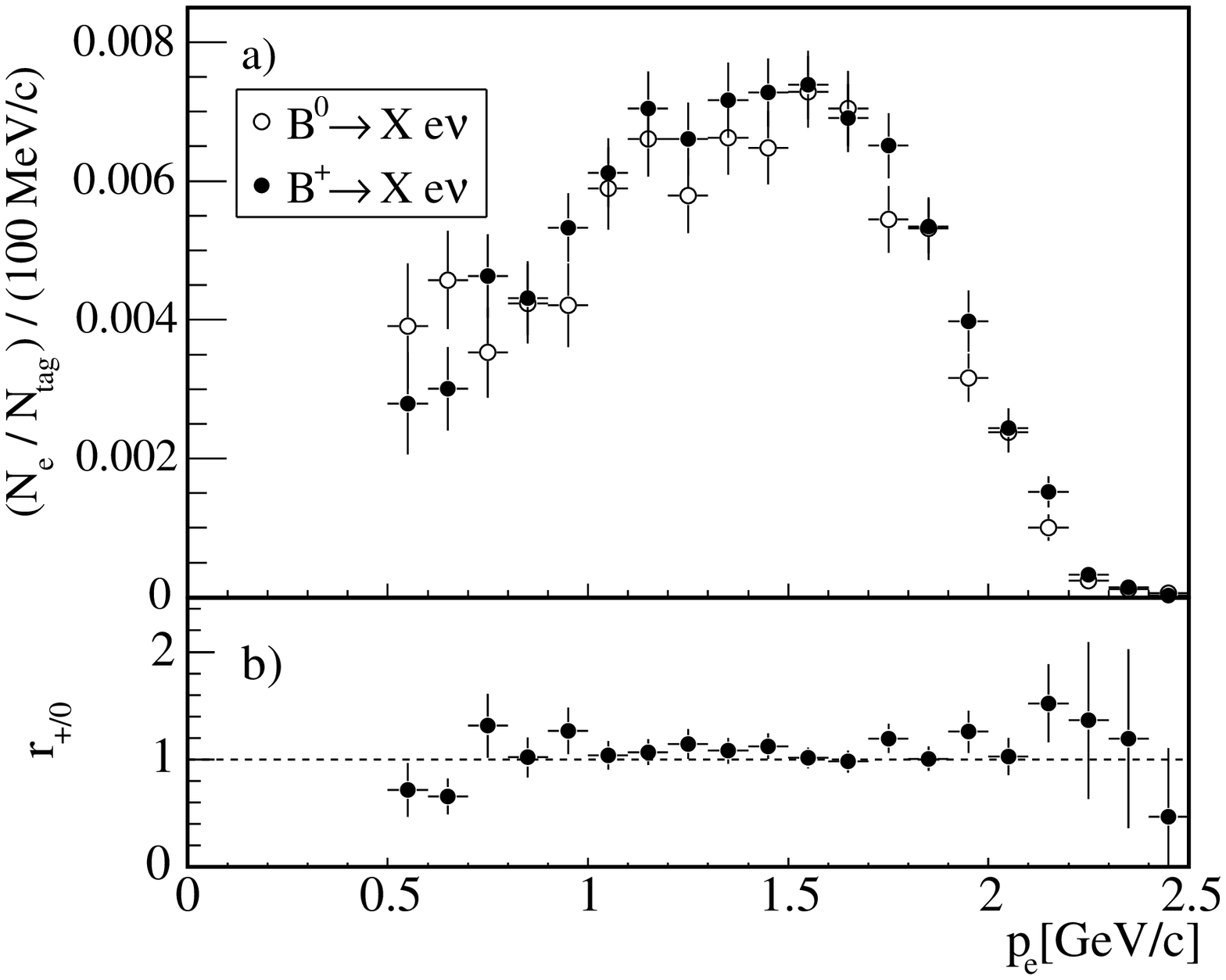} 
\caption{(a) Normalized momentum spectra of primary electrons after all efficiency corrections 
and (b) their ratio 
$r_{+/0}$ $= N_{\Bz} / N_{\Bu} \ \left(dN_{\Bu \to X e \nu}/dp\right) / \left(dN_{\Bz \to X e \nu}/dp\right)$.}
\label{finalspectra}
\end{center}
\end{figure}

After all corrections listed in Table~\ref{tblyields} have been applied, the 
inclusive momentum spectrum of electrons from semileptonic decays of 
\Bu mesons $dN_{\Bu \to X e\nu}/dp$ is given by the right-sign sample in 
\Bub-tagged events. Because of \BzBzb  oscillations, electrons from 
$\Bz \to X e \nu$ decays and $\Bz \to \Dbar X, \Dbar \to e^- \nue Y$ 
cascades contribute to both momentum spectra $dN^{\rm rs}_{\Bzb}/dp$ and 
$dN^{\rm ws}_{\Bzb}/dp$ of right- and wrong-sign samples in $\Bzb$ - tagged events, 
\begin{equation*}
\begin{aligned}
\frac{dN^{\rm rs}_{\Bzb}}{dp} &= \frac{dN_{\Bz \to X e\nu}}{dp}(1-\chi_m) + \frac{dN_{\Bz \to \Dbar \to X e\nu}}{dp}\chi_m \ , \\
\frac{dN^{\rm ws}_{\Bzb}}{dp} &= \frac{dN_{\Bz \to X e\nu}}{dp} \chi_m + \frac{dN_{\Bz \to \Dbar \to X e\nu}}{dp}(1-\chi_m) \ ,\\
\end{aligned}
\end{equation*}
with $\chi_m = (0.186 \pm 0.004)$~\cite{PDG} being the \BzBzb mixing parameter. 
We use these equations to determine the primary electron spectrum 
$dN_{\Bz \to X e\nu}/dp$ of neutral \B decays, which is shown in 
Fig.~\ref{finalspectra} together with $dN_{\Bu \to X e\nu}/dp$ after 
normalizations to the respective number of tags. 

We integrate these spectra between  $p_{\mathrm{min}}$ = 0.6 \gevc and 2.5 \gevc and apply
corrections for geometrical acceptance ($\epsilon_{\mathrm{geom}}=85\%$) and the small loss of electrons
due to bremsstrahlung in the detector material ($\epsilon_{\mathrm{brem}}=97.4 \pm 0.1 \%$)
to obtain the partial branching fractions $\hat{\BR}(\Bz \to X e \nu (\gamma))$ = $\BR(\Bz \to X e \nu (\gamma), p_e > p_{\mathrm{min}})$ for decays with any number of photons in the final state: 
\begin{multline*}
\hat{\BR}(\Bz \to X e \nu (\gamma))  = \brpartbna , \\
\hat{\BR}(\Bu \to X e \nu (\gamma))  = \brpartbca,  \\
\hat{\BR}(\B \to X e \nu (\gamma))  = \brpartba.
\end{multline*}
Table~\ref{tblsys} lists the contributions to the systematic errors. 
These results are in agreement with~\cite{belle,bbr1,bbr2}.
For the ratio of branching fractions,
$ \rpm(p_{\mathrm{min}}) = \BR(\Bu \to X e \nu (\gamma),p_e > p_{\mathrm{min}}) / \BR(\Bz \to X e \nu (\gamma),p_e > p_{\mathrm{min}})  $,
the result is  $ \rpm(0.6 \gevc) = 1.076 \pm 0.040_{\stat} \pm 0.029_{\syst}$.
For higher values of $p_{\mathrm{min}}$, the statistical error increases, 
while the systematic error decreases. At $p_{\mathrm{min}} = 1 \gevc$, the 
combined statistical and systematic error is minimal, leading to our
final result
$$ \rpm(1.0 \gevc) = 1.084 \pm 0.041_{\stat} \pm 0.025_{\syst}\ .$$ 

\begin{table}[!ht]
\caption{Breakdown of systematic errors on partial branching fractions $\hat{\BR}$ and
the ratio \rpm. Contributions in the upper part of this table are taken
to be uncorrelated for $\Bz$ and $\Bu$.}

\begin{tabular}{lcccc}
\hline 
\hline 
 &$\Delta \hat{\BR}^0 \newline [10^{-2}]$ &$\Delta \hat{\BR}^+ \newline [10^{-2}]$ &\multicolumn{2}{c}{$\Delta R_{+/0}  [10^{-2}] $}  \\
$p_{min} [\gevc]$ &0.6 &0.6 &0.6 &1.0  \\
\hline 
$N_{tags}$(uncorr.) &0.126 &0.141 &\ 0.020 &\ 0.020  \\
$\B_{tag}$ background &0.079 &0.122 &\ 0.014 &\ 0.012  \\
$\B \to D$ &0.079 &0.041 &\ 0.011 &\ 0.001  \\
$\B \to D_s$ &0.050 &0.054 &\ 0.008 &\ 0.002  \\
$\chi$ &0.037 & &\ 0.004 &\ 0.006  \\
$\dsxf$ &0.014 &0.064 &\ 0.004 &\ 0.003  \\
$\B \to \tau$ &0.019 &0.020 &\ 0.003 &\ 0.002  \\
\hline 
$N_{tags}$(corr.) &0.202 &0.253 &\ 0.004 &\ 0.004  \\
$e$ eff &0.134 &0.143 &\multicolumn{2}{c}{$<$0.001}  \\
 track eff. &0.085 &0.092 &\multicolumn{2}{c}{$<$0.001}  \\
 $D,D_s,\tau \to e$ &0.024 &0.020 &\multicolumn{2}{c}{$<$0.001}  \\
 conversion, Dalitz &0.025 &0.039 &\ 0.001 &$<$0.001  \\
faked $e$ &0.020 &0.027 &\multicolumn{2}{c}{$<$0.001}  \\
\hline 
\hline 
\end{tabular} 
\label{tblsys}
\end{table}

In summary, we have used electrons in \FourS\ decays tagged by a
fully reconstructed hadronic \B decay to measure the inclusive semileptonic
branching fractions of \Bz and \Bu mesons. The ratio of 
branching fractions, $ \rpm(1.0 \gevc) = 1.084 \pm 0.048$, is 
consistent with the ratio of \B lifetimes from direct measurements,
$\tau_{\Bu}/\tau_{\Bz}$ = $1.086 \pm 0.017$~\cite{PDG}. From this we conclude
that the semileptonic decay widths of charged and neutral \B mesons agree 
to a precision of 5\%, $\Gamma(\Bz \to X e \nu)$ /  $\Gamma(\Bu \to X e \nu)$ = 0.998 $\pm$ 0.047.

We are grateful for the excellent luminosity and machine conditions
provided by our \pep2\ colleagues, 
and for the substantial dedicated effort from
the computing organizations that support \babar.
The collaborating institutions wish to thank 
SLAC for its support and kind hospitality. 
This work is supported by
DOE
and NSF (USA),
NSERC (Canada),
IHEP (China),
CEA and
CNRS-IN2P3
(France),
BMBF and DFG
(Germany),
INFN (Italy),
FOM (The Netherlands),
NFR (Norway),
MIST (Russia), and
PPARC (United Kingdom). 
Individuals have received support from the 
A.~P.~Sloan Foundation, 
Research Corporation,
and Alexander von Humboldt Foundation.


\newpage


\begin{thebibliography}{99}
\bibitem{conjugate}
Charge conjugation is implied throughout the paper.
\end{thebibliography}
\end{document}